\documentclass[conference]{IEEEtran}
\usepackage{amsmath,amssymb,multicol,graphicx,gensymb,flushend,booktabs,array,url,tikz}
\IEEEoverridecommandlockouts

\begin{document}

\title{A Clinical--Metabolic Approach to Determining Calcium Requirements: Interplay of the Glucose Axis, PTH--Vitamin D Signalling and Bone Mineral Density}

\author{\IEEEauthorblockN{Alex V\'elez Cruz\IEEEauthorrefmark{1},
Oscar Dam\IEEEauthorrefmark{2}, \& Jairo Rond\'on\IEEEauthorrefmark{3}}
\IEEEauthorblockA{\IEEEauthorrefmark{1}Biomedical Engineering Department, Polytechnic University of Puerto Rico, San Juan, Puerto Rico, USA\\
alvelez@pupr.edu}
\IEEEauthorblockA{\IEEEauthorrefmark{2}Universidad Nacional Polit\'ecnica Antonio Jos\'e de Sucre (UNEXPO), Puerto Ordaz, Venezuela\\
odamg@unexpo.edu.ve}
\IEEEauthorblockA{\IEEEauthorrefmark{3}Biomedical \& Chemical Engineering Departments, Polytechnic University of Puerto Rico, San Juan, Puerto Rico, USA\\
jrondon@pupr.edu}
}

\maketitle

\begin{abstract}
Calcium homeostasis is maintained by an endocrine network comprising parathyroid hormone (PTH), calcitriol and calcitonin, which defends serum calcium at the expense of skeletal reserves. Serum calcium is therefore an unreliable indicator of nutritional calcium status, and empirical supplementation remains common. This narrative review proposes a clinical--metabolic approach to estimating individual calcium requirements that integrates bone mineral density (BMD), vitamin D status and the compensatory PTH response. We make three contributions. First, we present a sigmoidal phosphate--PTH model in which a glucose-dependent gain term ($\Delta k_{\mathrm{gluc}}$) captures the heightened parathyroid sensitivity associated with dysglycaemia, and we illustrate it with simulated values for healthy adults, women and individuals with gluco-osteoporosis. Second, we translate the model into a four-step decision algorithm that frames the intake gap as a net balance of intestinal absorption, renal excretion and bone turnover. Third, we incorporate the osteocyte--osteoblast bioenergetic axis (sclerostin/Wnt signalling, GLUT1-mediated glucose uptake, aerobic glycolysis and osteocalcin--insulin crosstalk) to explain why calcium supplementation may be ineffective when cellular energy supply or insulin sensitivity is impaired. The framework aims to prevent secondary hyperparathyroidism, avoid over supplementation and preserve skeletal integrity. The model values are simulated and require prospective clinical validation.
\end{abstract}


\section{Introduction}
Calcium is the most abundant mineral in the human body. Approximately 99\% is stored in the skeletal matrix as hydroxyapatite crystals, while the small circulating fraction is vital for cell signalling, coagulation and neuromuscular function \cite{r1}. This distribution creates what we term the ``serum calcium paradox'': because the organism prioritizes hemodynamic and neuromuscular stability, normal circulating calcium does not guarantee an adequate intake. When supply is insufficient, the deficit is silently drawn from bone, and the skeleton, rather than the blood, records the cost of nutritional inadequacy.

Current guidance for calcium intake relies largely on population-level recommended dietary allowances stratified by age and sex \cite{r1,r2}. Although these reference values are essential for public health, they do not account for individual differences in intestinal absorption, vitamin D status, parathyroid responsiveness or the metabolic milieu in which bone cells operate. In particular, the contribution of glucose metabolism to parathyroid sensitivity and to osteoblast energetics has rarely been incorporated into practical intake calculations. As a result, clinicians frequently resort to empirical supplementation that may be insufficient in some patients and excessive in others.

The aim of this review is to bridge this gap by proposing a clinical--metabolic framework for estimating individual calcium requirements. The principal contributions are threefold: (i) a glucose-modulated phosphate--PTH model that quantifies how dysglycaemia shifts parathyroid sensitivity; (ii) a four-step clinical algorithm that converts BMD, vitamin D and PTH data into an individualized intake target; and (iii) a bioenergetic perspective on the osteocyte--osteoblast axis that explains why calcium repletion alone may fail in energy-deficient or insulin-resistant states. The remainder of the article is organized accordingly: Section II summarizes the physiology of the PTH--vitamin D axis and presents the model, Section III addresses the assessment of bone status, Section IV describes the algorithm, Section V reviews bone bioenergetics, and Sections VI and VII discuss clinical implications, limitations and future work.

\section{Physiology of Calcium Homeostasis and the PTH--Vitamin D Axis}
\subsection{Parathyroid Response to a Reduced Calcium Supply}
When calcium intake falls, the transient decrease in ionized calcium activates calcium-sensing receptors (CaSR) on parathyroid chief cells and increases PTH secretion. PTH then acts through three principal mechanisms: (i) bone resorption, whereby osteoclast activity is stimulated to release calcium from the skeleton; (ii) activation of vitamin D, through induction of renal 1$\alpha$-hydroxylase, which converts 25-hydroxyvitamin D into its active form, calcitriol; and (iii) renal reabsorption, which reduces urinary calcium loss. Calcitriol, in turn, enhances intestinal calcium absorption \cite{r3}. If this compensatory loop is sustained by a chronically inadequate supply, bone resorption exceeds formation, and physiological remodeling becomes pathological.

\subsection{Phosphate--PTH Relationship and Glucose Modulation}
Serum phosphate (Pi) is a further determinant of PTH secretion, because phosphate retention stimulates the parathyroid gland both directly and by lowering ionized calcium and calcitriol. We therefore used serum phosphate as a physiological input to model the PTH response and examined how glycaemic status modifies this relationship. The model coordinates, generated by simulation, are shown in Table~\ref{tab:modelcoords}.

\begin{table*}[!t]
\caption{Model coordinates (simulated values) for the phosphate--PTH relationship under three metabolic profiles. The $k$ is the dimensionless gain (sensitivity) factor of (1). Source: Osteoawareness (private dataset provided by authors).}
\label{tab:modelcoords}
\centering
\small
\begin{tabular}{@{}llll@{}}
\toprule
\textbf{Pi (mg/dL)} & \textbf{Healthy ($k=1.00$)} & \textbf{Woman ($k=1.35$)} & \textbf{Gluco-osteoporosis ($k=1.80$)} \\
\midrule
3.5 (low) & 20 pg/mL & 15 pg/mL & 38 pg/mL (already elevated) \\
4.0 (borderline) & 30 pg/mL & 35 pg/mL & 65 pg/mL (upper reference limit) \\
4.2 (threshold, $Pi_0$) & 40 pg/mL & 55 pg/mL & 90 pg/mL (clearly elevated) \\
4.8 (critical) & 65 pg/mL & 95 pg/mL & 125 pg/mL (severe elevation) \\
\bottomrule
\end{tabular}
\end{table*}

The calcium--parathyroid response is described by a sigmoidal (logistic) function in which the glucose-dependent term modifies the gain of the system:
\begin{equation}
\mathrm{PTH}(Pi)=\frac{L}{1+e^{-[(k+\Delta k_{\mathrm{gluc}})(Pi-Pi_0)]}}+B_{\mathrm{gluc}}.
\label{eq:pth}
\end{equation}
where $L$ is the maximal secretory capacity of the parathyroid glands; $B_{\mathrm{gluc}}$ is the basal PTH level (minimum constant secretion), adjusted for glycaemic status; $Pi_0$ is the inflection point, i.e. the metabolic threshold at which the gland mounts a rapid response (around 4.2 mg/dL); $k$ is the proportionality (gain) factor, which reflects system sensitivity and varies with sex and metabolic profile; and $\Delta k_{\mathrm{gluc}}$ is the increase in parathyroid sensitivity attributable to impaired glucose metabolism, whereby the gland responds disproportionately to a given phosphate load. For example, if $k=1.35$ in a woman without metabolic compromise, the effective gain in gluco-osteoporosis may rise to between 1.6 and 1.8.

\begin{figure*}[!t]
    \centering
    \includegraphics[width=0.85\textwidth]{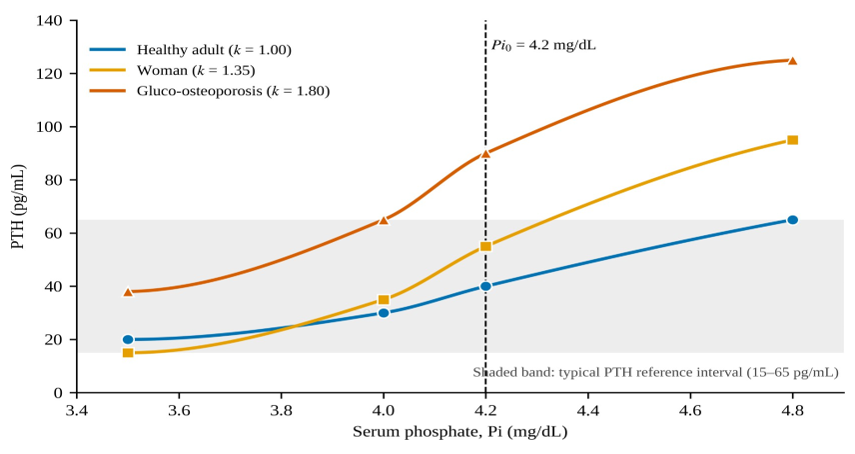}
    \caption{Effect of glycaemic status on the phosphate--PTH relationship under three simulated metabolic profiles (healthy adult, woman, and gluco-osteoporosis; see Table~I). Markers show the model coordinates; curves are smooth monotone interpolations between them. The dashed line marks the inflection threshold ($P_{i0}=4.2$ mg/dL), and the shaded band represents the typical PTH reference interval (15--65 pg/mL).}
    \label{fig:pth_phosphate}
\end{figure*}

Throughout this article, ``gluco-osteoporosis'' denotes osteoporosis arising in the context of impaired glucose metabolism (diabetes mellitus or chronic hyperglycaemia). As Fig.~\ref{fig:pth_phosphate} shows, the combined effect of glycaemia and phosphate on PTH is substantial: at a phosphate concentration of 4.2 mg/dL, the simulated PTH is 40 pg/mL in the healthy profile but 90 pg/mL in gluco-osteoporosis, and at 3.5 mg/dL the latter already approaches twice the healthy value. These simulated trajectories suggest that individuals with dysglycaemia enter the compensatory PTH range at lower phosphate levels, and therefore at an earlier stage of calcium insufficiency, than their metabolically healthy counterparts.

\section{Assessment of Bone Status by Densitometry}
Skeletal depletion is quantified by dual-energy X-ray absorptiometry (DXA). The T-score expresses BMD as the number of standard deviations (SD) from the mean peak bone mass of a young adult reference population. According to the World Health Organization criteria, a T-score of $-1.0$ to $-2.5$ SD indicates low bone mass (osteopenia), and a T-score $\leq -2.5$ SD indicates osteoporosis \cite{r4}. A reduced T-score signals a depleted calcium reserve and, in the context of an elevated PTH, justifies a more assertive intake strategy to arrest PTH-mediated resorption. However, because BMD does not capture bone quality, particularly in diabetes (Section VI), densitometric data should be interpreted together with biochemical and clinical risk factors.

\section{Proposed Methodology for Determining Calcium Intake}
To determine the individual requirement, we propose a four-step protocol, summarized in Table~\ref{tab:algorithm} and described below.

\begin{table*}[!t]
\caption{Four-step clinical algorithm for estimating individual calcium requirements.}
\label{tab:algorithm}
\centering
\small
\begin{tabular}{@{}p{0.05\textwidth}p{0.20\textwidth}p{0.67\textwidth}@{}}
\toprule
\textbf{Step} & \textbf{Action} & \textbf{Output / decision rule} \\
\midrule
1 & Quantify baseline intake & 24-h dietary recall or food-frequency questionnaire $\rightarrow I_{\mathrm{diet}}$ (mg/day) \\
2 & Analyze biomarkers & Serum calcium; 25(OH)D; intact PTH; phosphate; glycaemic status; renal function. PTH above the upper limit suggests negative calcium balance \\
3 & Calculate nutritional deficit & Supplement requirement $S=\max(0,\mathrm{RDA}-I_{\mathrm{diet}})$ (2), adjusted for age, sex and BMD \\
4 & Apply vitamin D correction & If 25(OH)D is deficient, administer a loading dose before or together with calcium, then reassess PTH and 25(OH)D \\
\bottomrule
\end{tabular}
\end{table*}

\subsection{Quantification of Baseline Intake}
Habitual intake is estimated by a 24-hour dietary recall or a food-frequency questionnaire, focusing on dairy products, cruciferous vegetables and fortified foods \cite{r2}. The resulting value ($I_{\mathrm{diet}}$) is the starting point of the calculation.

\subsection{Biomarker Analysis}
\begin{itemize}
\item \textbf{Serum calcium:} useful only for detecting acute disorders or severe malabsorption, because it is buffered by skeletal release.
\item \textbf{25-hydroxyvitamin D [25(OH)D]:} levels $\geq 30$ ng/mL (75 nmol/L) are considered optimal for calcium absorption \cite{r5}.
\item \textbf{PTH:} values above the upper reference limit suggest a negative calcium balance \cite{r3}.
\end{itemize}

\subsection{Calculation of the Nutritional Deficit}
The recommended intake (RDA) depends on age and sex; for most adults it is 1000 mg/day, rising to 1200 mg/day in women older than 50 years and in all adults older than 70 years \cite{r1}. The supplementation requirement \cite{r6} is then calculated as the shortfall between the recommended and the habitual dietary intake:
\begin{equation}
S=\max\left[0,\left(\mathrm{RDA}-I_{\mathrm{diet}}\right)\right].
\label{eq:supplement}
\end{equation}
where $S$ is the supplemental calcium dose (mg/day). Conceptually, the intake must also satisfy the net calcium balance, in which $\alpha$ is the fractional intestinal absorption, $I$ is the total intake, and $E_{\mathrm{ren}}$ and $E_{\mathrm{other}}$ are renal and other obligatory losses, respectively. A positive balance ($B_{\mathrm{Ca}}>0$) indicates net skeletal deposition, whereas a negative balance indicates net bone loss.
\begin{equation}
B_{\mathrm{Ca}}=\alpha I-\left(E_{\mathrm{ren}}+E_{\mathrm{other}}\right).
\label{eq:balance}
\end{equation}

\subsection{Vitamin D Correction Factor}
When vitamin D is deficient (25(OH)D $<20$ ng/mL) \cite{r1,r5}, intestinal calcium absorption declines markedly (in some estimates from about 30\% to less than 15\%). In this scenario, the calculated intake should be accompanied by a loading dose of vitamin D to restore calcium absorption before the benefit of supplementation can be expected.

\section{Bioenergetic Dynamics and Metabolic Coupling: Osteocyte--Osteoblast Axis}
Bone repair and remodelling depend on the efficient transfer of signals and energy. Bone has evolved a system of metabolic coupling in which the osteocyte acts as the ``sensor and director'' and the osteoblast as the ``high-demand worker''. This section describes the cellular mechanisms through which calcium availability, glucose supply and endocrine signalling converge.

\subsection{The Osteocyte as a Metabolic Sensor and Flow Regulator}
Osteocytes, embedded in the mineralised matrix, form a functional network through the lacunocanalicular system. Their principal role is not matrix synthesis but mechanotransduction \cite{r7}. Two mechanisms are particularly relevant:
\begin{itemize}
\item \textbf{Detection of microdamage.} Under wear or mechanical stress, osteocytes sense deformation of the interstitial fluid, which increases extracellular ATP production; ATP then acts as a signalling molecule.
\item \textbf{Sclerostin-mediated regulation.} Under normal loading, or when repair is required, osteocytes reduce sclerostin expression. Sclerostin is an inhibitor of the Wnt/$\beta$-catenin pathway; its downregulation releases this pathway and triggers glycolytic metabolism in osteoblasts.
\end{itemize}

\subsection{Osteoblasts and the Warburg-Like Effect on Bone Formation}
Osteoblasts have a massive energy demand owing to type I collagen synthesis and the secretion of matrix vesicles for mineralisation. Notably, despite the presence of oxygen, osteoblasts preferentially use aerobic glycolysis, analogous to the Warburg effect described in tumour cells, to generate energy rapidly during the formation phase \cite{r8,r9}. Two aspects are relevant to calcium requirements:
\begin{itemize}
\item \textbf{Glucose dependence.} Glucose uptake by osteoblasts is mediated mainly by the transporter GLUT1. Evidence indicates that PTH and Wnt signalling markedly increase GLUT1 expression and glucose uptake \cite{r8,r9,r10}.
\item \textbf{Relationship with calcium.} Calcium transport towards the mineralisation front is an ATP-dependent process involving the Ca$^{2+}$-ATPase pump. Without a steady supply of glucose and ATP, the osteoblast cannot precipitate hydroxyapatite crystals, regardless of the serum calcium concentration.
\end{itemize}

\subsection{Energetic Synergy During Repair}
During the repair of a microfracture or physiological wear, the following bioenergetic sequence is proposed:
\begin{itemize}
\item \textbf{Local hypoxia and signalling.} Damage interrupts microscopic blood flow. Sensor osteocytes secrete RANKL and reduce osteoprotegerin (OPG) to recruit osteoclasts that clear the damaged area \cite{r7}.
\item \textbf{Lactate transfer.} Recent studies suggest a ``shuttle'' model in which osteocytes may metabolise glucose and transfer intermediates, such as lactate, to osteoblasts to support their biosynthesis \cite{r9,r10}. This model remains hypothetical and requires further confirmation.
\item \textbf{Energetic cost of collagen.} Procollagen synthesis requires hydroxylation of proline and lysine residues, a process dependent on ascorbate, oxygen and iron, and consumes a significant fraction of the cellular energy budget.
\end{itemize}

\subsection{Endocrine Interaction: PTH, Insulin and Osteocalcin}
The energetic relationship is closed by an endocrine loop. PTH stimulates glucose uptake in osteoblasts, providing the energy required for bone formation. Osteoblasts, in turn, secrete osteocalcin in its undercarboxylated form, which acts on the pancreas to increase insulin secretion and on adipose tissue to improve glucose sensitivity \cite{r11}. Thus, bone appears to regulate its own fuel supply so that repair processes are energetically supported, although the magnitude of this loop in humans remains under investigation.

\section{Clinical Implications of Bone Bioenergetics}
When calculating calcium intake, the clinician should consider two metabolic scenarios in which calcium repletion alone is unlikely to be sufficient:
\begin{itemize}
\item \textbf{Energy malnutrition.} In patients with a severe caloric deficit, calcium supplementation may be less effective, as the osteoblast may enter a quiescent state owing to ATP deficiency.
\item \textbf{Diabetes and bone repair.} Insulin resistance impairs glucose uptake by osteoblasts. This may explain why patients with diabetes often show a deceptively normal or even elevated BMD but poor bone quality and a high fracture risk \cite{r12}.
\end{itemize}

\section{Discussion}
\subsection{The Risk of Oversupplementation}
``More'' is not necessarily ``better''. Intakes above 2000--2500 mg/day \cite{r1} have been associated with kidney stones \cite{r13} and, in some controversial studies, with vascular calcification and cardiovascular events \cite{r14}. Dietary sources should therefore always take priority, with supplements (carbonate or citrate) reserved exclusively to cover the calculated gap.

\subsection{Strengths, Limitations and Future Directions}
The principal strength of the proposed framework is its integration of three layers that are usually considered separately: the endocrine response (PTH--vitamin D--phosphate), the structural status of the skeleton (BMD) and the cellular energy supply (glucose--ATP). This integration provides a mechanistic rationale for individualizing intake and for anticipating non-response to supplementation.

Several limitations must be acknowledged. First, the values in Table~\ref{tab:modelcoords} and Fig.~\ref{fig:pth_phosphate} are simulated and derive from a private dataset; they illustrate the behavior of the model rather than constitute empirical evidence. Second, the gain factors ($k$) and the glucose-dependent increment ($\Delta k_{\mathrm{gluc}}$) have not been estimated from patient-level data, and their dependence on age, renal function and medication has not been explored. Third, the bioenergetic mechanisms described in Section V derive largely from cellular and animal studies, and their translation to clinical practice is still incomplete. Future work should (i) calibrate the model against prospectively collected phosphate, PTH, 25(OH)D and glycaemic data; (ii) evaluate whether the algorithm improves BMD, fracture or biochemical outcomes compared with standard supplementation; and (iii) define phosphate and PTH thresholds in specific populations, including patients with diabetes and chronic kidney disease.

\section{Conclusions}
Determining calcium requirements is a dynamic process. A normal serum level is not sufficient; it is imperative to evaluate the ``health of the store'' (BMD) and the ``state of the transporter'' (vitamin D), together with the parathyroid response and the metabolic context in which bone cell's function. By coupling a glucose-modulated phosphate--PTH model with a stepwise intake algorithm and a bioenergetic interpretation of osteocyte--osteoblast activity, the proposed framework supports personalized intervention that not only prevents fractures but also optimizes overall metabolic function. Prospective validation is the next necessary step.

\end{document}